\documentclass[12pt]{article}
\usepackage{graphicx}
\usepackage{amssymb}
\usepackage{epsfig}
\usepackage{amsmath}
\newcommand{\be}{\begin{equation}}
\newcommand{\ee}{\end{equation}}

\newcommand{\LA}{\langle}
\newcommand{\RA}{\rangle}

\newcommand{\bea}{\begin{eqnarray}}
\newcommand{\eea}{\end{eqnarray}}

\newcommand{\p}{\partial}

\newcommand{\ie}{{\it i.e.}}

\newcommand{\m}{\mu}\newcommand{\n}{\nu}
\newcommand{\D}{\Delta}
\newcommand{\ta}{\theta}
\newcommand{\f}{\phi}
\newcommand{\K}{\kappa}

\DeclareSymbolFont{AMSa}{U}{msa}{m}{n}
\DeclareSymbolFont{AMSb}{U}{msb}{m}{n}
\DeclareMathSymbol{\fieldR}{\mathalpha}{AMSb}{"52}

\newif\ifpdf
\ifx\pdfoutput\uundefined
\pdffalse 
\else
\pdfoutput=1 
\pdftrue
\fi

\begin{document}
\begin{titlepage}
\begin{center}
\hfill {\tt YITP-SB-06-33}\\
\hfill {\tt hep-th/0608077}\\
\vskip 20mm


{\Large {\bf Operator Product Expansion of Higher Rank \\ Wilson Loops 
 from D-branes and Matrix Models}}

\vskip 10mm
\renewcommand{\thefootnote}{\fnsymbol{footnote}}

{\bf Simone Giombi, Riccardo Ricci, and Diego Trancanelli
\footnote{E-mails: {\tt sgiombi, rricci,
dtrancan@max2.physics.sunysb.edu.}}}

\vskip 4mm {\em  C. N. Yang Institute for Theoretical
Physics,}\\
{\em State University of New York at Stony Brook}\\
{\em Stony Brook, NY 11794-3840, USA}\\
[2mm]

\end{center}

\vskip 1in

\renewcommand{\thefootnote}{\arabic{footnote}}
\setcounter{footnote}{0}

\begin{center} {\bf ABSTRACT }\end{center}
\begin{quotation}
\noindent
In this paper we study correlation functions of circular Wilson loops in higher dimensional representations with chiral primary operators of $\mathcal{N}=4$ super Yang-Mills theory. 
This is done using the recently established relation between higher rank Wilson loops in gauge theory and D-branes with electric fluxes in supergravity.
We verify our results with a matrix model computation, finding perfect agreement in both the symmetric and the antisymmetric case.

\end{quotation}

\vfill
\flushleft{August 11, 2006}

\end{titlepage}

\eject \tableofcontents
\section{Introduction}
\setcounter{equation}{0}
In the AdS/CFT correspondence local gauge invariant operators
are matched with bulk supergravity fields evaluated at the boundary
of the AdS space \cite{Maldacena:1997re}\cite{Gubser:1998bc}\cite{Witten:1998qj}.
An important role in the correspondence is also played by non local gauge invariant operators,
the most notable example being the Wilson loop.

In $\mathcal{N}=4$
super Yang-Mills theory Wilson loops are defined (in Euclidean signature) as
\bea
W_{\mathcal{R}}(\mathcal{C})=\frac{1}{\mbox{dim}\mathcal{R}}\mbox{Tr}_{\mathcal{R}}\mathcal{P}
\exp\oint_{\mathcal{C}}d\tau\left[iA_{\m}(\tau)\dot{x}^{\m}+\Phi_{I}
(\tau)\theta^I|\dot{x}|\,\right]\label{wilson0}\eea where $A_\m$ is
the gauge field and $\Phi_I$ are the six scalars of the
$\mathcal{N}=4$ multiplet, and $\theta^I$ is a constant unit vector
in $\mathbb{R}^6$. The data which characterize the Wilson loop are
the shape of the integration contour $\mathcal{C}$ and the
representation $\mathcal{R}$ of the gauge group. Supersymmetry
restricts $\mathcal{C}$ to be a straight line or a circle \cite{Zarembo:2002an} 
\footnote{In this paper we consider only half-BPS operators. For loops preserving less supersymmetry see \cite{Zarembo:2002an}\cite{Drukker:2005cu}\cite{Drukker:2006ga}.}, 
while
$\mathcal{R}$ may be arbitrary. It is a well-established part of the
AdS/CFT correspondence that Wilson loops in
the fundamental representation are associated with a classical string
surface of minimal area landing on the loop
\cite{Rey:1998ik}\cite{Maldacena:1998im}.
The worldsheet area,
being the string infinitely long, is formally infinite but the string
action can nevertheless be made finite by adding suitable
counterterms \cite{Drukker:1999zq}. The expectation value of the Wilson loop is then the
partition function of the regularized string associated to it.

Recently circular Wilson loops in representations other than the fundamental
have been a very active field of investigation. A holographic
dictionary has emerged, where probe branes in the bulk
are related to higher rank Wilson loops in the
boundary
\cite{Drukker:2005kx}\cite{Yamaguchi:2006tq} \cite{Gomis:2006sb}\cite{Rodriguez-Gomez:2006zz} 
\footnote{For 't Hooft loops see \cite{Drukker:2005kx}\cite{Chen:2006iu}, and for applications to finite temperature see \cite{Hartnoll:2006hr}.}.
In particular, D3 branes with $AdS_2\times S^2$ worldvolume and $k$
units of fundamental string charge dissolved on them have been
proved to compute expectation values of Wilson loops in the rank $k$
symmetric representation \footnote{The brane
probe approximation breaks down for $k$ much larger than $N$. In this limit the branes backreact deforming the geometry
into the supergravity solutions studied in \cite{Yamaguchi:2006te}
and \cite{Lunin}.}.
This picture is the natural
generalization to the $AdS_5\times S^5$ background of the
idea that a fundamental string ending on a D3 brane in flat space
can be described in terms of a curved D3 brane with a localized spike
carrying a unit of
electric flux, as first proposed by Callan and Maldacena \cite{Callan:1997kz}.
On the other hand, D5 branes with
$AdS_2\times S^4$ worldvolume and $k$ units of string charge
correspond to Wilson loops in the rank $k$ antisymmetric
representation. Both these branes are half-BPS and
preserve the same isometries of (\ref{wilson0}), namely $SO(2,1)\times
SO(3)\times SO(5)$. They pinch off at the boundary of
$AdS_5$ landing on the curve that defines the Wilson loop.

The intuitive reason for considering these objects is that,
to build a Wilson loop in the rank $k$
representation, one would start with considering $k$ coincident fundamental strings.
The D$3_k$ and D$5_k$ branes can then be thought of as coming from an
Emparan-Myers polarization effect \cite{Emparan:1997rt}\cite{Myers:1999ps},
which, for $k$ sufficiently
large, blows up a $S^2\subset AdS_5$ or a $S^4\subset S^5$
from the worldsheet of the $k$ coincident strings.
This is
reminiscent of the interpretation of gravitons with large momenta
as D3 branes wrapping a $S^3\subset S^5$ (giant gravitons) or a $S^3\subset AdS_5$ (dual giant gravitons). We can then regard the
D$3_k$ and D$5_k$ branes as dual giant and giant Wilson loops, respectively.

This brane picture has the advantage of automatically encoding the interactions between the coincident strings \cite{Gross:1998gk} and yields all non planar contributions to the expectation value of the higher rank Wilson loop \cite{Drukker:2005kx}.

It is well-known that the expectation value of circular Wilson loops in the fundamental representation
can be computed with a quadratic Hermitian matrix model \cite{Erickson:2000af}\cite{Drukker:2000rr}. It has been conjectured that this can be extended to
higher rank loops and matrix model computations have provided a successful check of the holographic dictionary just discussed
\cite{Drukker:2005kx}\cite{Okuyama:2006jc}\cite{Hartnoll:2006is}.

A small circular Wilson loop, when probed from a
distance much larger than its characteristic size, can be expanded
in a series of local operators of different conformal dimension
\cite{Berenstein:1998ij}.
The operators which
are allowed to appear in the expansions must preserve the same
symmetries of (\ref{wilson0}) and therefore must be
bosonic, gauge invariant and $SO(5)$ invariant.
The conformal dimension of some of these operators is not
protected by the superconformal algebra and therefore
they receive large anomalous dimensions and decouple in the strong coupling regime.
An important class of operators which have protected dimensions and appear in the operator product expansion are the chiral primary operators.
The correlator with a local operator can then be read off from the expansion of the Wilson loop
 \cite{Berenstein:1998ij} \footnote{For a nice review see \cite{Semenoff:2002kk}.}.

In this paper, we use the D$3_{k}$ and D$5_{k}$ branes to compute
the correlation function between a circular Wilson loop in a higher
representation and a chiral primary operator in the fundamental representation.
We do this by studying the coupling to the brane worldvolume
of the supergravity modes dual to the chiral primaries.
These modes propagate from the insertion of the local operator on the boundary to the brane worldvolume in the bulk.

The paper is organized as follows.
In sections 2 and 3 we review the operator product expansion of the Wilson loop and how to compute correlation functions when the Wilson loop is described in terms of a fundamental string worldsheet. To evaluate this one needs to study the harmonic expansion on $S^5$ of the bulk fields which couple to the worldsheet.

Following the philosophy outlined before, we then replace
the fundamental string with the D$3_k$ and D$5_k$
branes. We start by investigating the symmetric case in section 4. We
expand the brane action to linear order in the fluctuations of the bulk fields and find how it couples to the relevant
 supergravity modes. Using the procedure reviewed in
section 3 we compute the correlation function.
In the limit of small $k$ we recover the
previously known result derived using the fundamental string.

We then move on to the analysis of the antisymmetric case. The
D$5_{k}$ brane now extends also in the $S^5$ directions. Also in
this case we compute the correlator between the Wilson loop and a
chiral primary operator
 and check
that it yields the correct string limit.

As a further check, in section 5, we compare our results against the
expressions coming from the normal matrix model introduced in this context in
\cite{Okuyama:2006jc} and find perfect agreement both in the
symmetric and antisymmetric case.

In the appendix we collect some facts about spherical harmonics and
orthogonal polynomials that we have used in the paper.

\section{Kaluza-Klein expansion}
\setcounter{equation}{0}

In this section we review the expansion in spherical harmonics for
type IIB supergravity on $AdS_5\times S^5$ \cite{Kim:1985ez}, and identify
the bulk excitations associated to turning on a chiral
primary operator in the dual ${\cal N}=4$ gauge
theory \cite{Lee:1998bx}. These will be later used to construct the
coupling of the various supergravity modes to the D$3_k$ and D$5_k$
branes.

The Einstein equations read \footnote{In our conventions Latin
indices run over the whole 10 dimensional manifold while Greek
indices $\mu,\nu,\ldots$ and $\alpha,\beta,\ldots$ run over $AdS_5$ and
$S^5$ respectively. We also choose units in which 
$R_{AdS_5}=R_{S^5}=1$.} \be R_{mn}=\frac{1}{96}
F_{mijkl}F_n^{\,\,\,ijkl} \ee where the 5-form field strength
$F_{(5)}$ is self-dual. In the Poincare patch, the $AdS_5\times
S^5$ solution reads
\begin{eqnarray}
&&ds^2 = \frac{1}{z^2} \left(dz^2+d\vec{x}^{\,2}\right)
+ d\Omega_5^2\\
&& \bar{F}_{{\m_1}{\m_2}{\m_3}{\m_4}{\m_5}}=
-4\epsilon_{{\m_1}{\m_2}{\m_3}{\m_4}{\m_5}},~~~~~
\bar{F}_{{\alpha_1}{\alpha_2}{\alpha_3}{\alpha_4}{\alpha_5}} =
-4\epsilon_{{\alpha_1}{\alpha_2}{\alpha_3}{\alpha_4}{\alpha_5}}.
\end{eqnarray}
 The
fluctuations around the background geometry can be parametrized as
follows
\begin{eqnarray}
G_{mn}&=& g_{mn} + h_{mn}\\
 h_{\alpha\beta}&=&h_{(\alpha\beta)}+\frac{h_2}{5}g_{\alpha\beta},~~~~~~  g^{\alpha\beta}h_{(\alpha\beta)}=0 \label{halphabeta}\\
h_{\mu\nu} &=& h'_{\mu\nu} -{h_2 \over 3} g_{\mu\nu}, ~~~~~~~~
g^{\mu\nu}h'_{(\mu\nu)} = 0\\ F&=&\bar{F} + \delta F, ~~~~~~~~~~~~~~ \delta
F_{ijklm}=
5\nabla_{[i}a_{jklm]}
\end{eqnarray}
where $h_2$ is the trace of the metric on the five-sphere,
$h_2\equiv h_{\alpha\beta}g^{\alpha\beta}$. Note that the fields
$h_{\mu\nu}$ and $h'_{\mu\nu}$ are related by a $d=5$ Weyl shift.
To identify the bulk excitation in $AdS_5$ we expand the
fluctuations as follows \footnote{We do not consider the harmonic
expansion of $h_{(\alpha\beta)}$ as this fluctuation is related to
$Q^2\bar{Q}^2$ descendants of chiral primaries in the dual super
Yang-Mills theory.}
\begin{eqnarray}
{h'}_{\m\n} &=& \sum  {h'}^I_{\mu\nu}(x)Y^{I}(y)\\
 h_2 &=& \sum h_2^{I}(x) Y^{I}(y) \\
 a_{\mu_1\mu_2\mu_3\mu_4} &=& \sum
 a^{I}_{\mu_1\mu_2\mu_3\mu_4}(x)Y^{I}(y)\\
 a_{\alpha_1\alpha_2\alpha_3\alpha_4} &=&-4
\sum\epsilon_{\alpha\alpha_1\alpha_2\alpha_3\alpha_4}b^{I}(x)\nabla^{\alpha}Y^{I}(y)
\end{eqnarray}
where $x$ and $y$ refer to the $AdS_5$ and $S^5$ coordinates
respectively, and $Y^I$ are scalar spherical harmonics on the
5-sphere which satisfy \footnote{We include a brief review of spherical harmonics in the appendix.} \be
\nabla^{\alpha}\nabla_{\alpha}Y^I=-\D(\D+4)Y^I. \ee Spherical
harmonics on $S^5$ can be classified in terms of the $SO(6)\simeq
SU(4)$ R-symmetry group. In particular scalar harmonics belong to
the $[0,\D,0]$ representation. The fields $h_2$ and $b$ appear
coupled in the linearized equation of motions. Their equations can
be diagonalized introducing  the linear combinations
\cite{Lee:1998bx}
\begin{eqnarray}
&s^I &={1\over 20 (\D+2)}[h_2^I-10(\D+4)b^I]\\
&t^I &={1\over 20 (\D+2)}[h_2^I+10\D{b^I}]\label{t}
\end{eqnarray}
which obey the equations of motion
\begin{eqnarray}\nabla_{\mu}\nabla^{\mu}\,s^I &=& \D(\D-4)\,
s^I\\
\nabla_{\mu}\nabla^{\mu}\,t^I &=& (\D+4)(\D+8) \,t^I.
\end{eqnarray}

A scalar field in AdS with $m^2=\D(\D-4)$ (with $\D\ge 2$) transforming in the
$[0,\D,0]$ representation corresponds to a chiral primary operator
$ {\cal O}_{\D}$ of conformal dimension $\D$. Therefore, to linear
order,
the scalar field $s^I$ corresponds to chiral primaries
in the dual gauge theory. On the other hand, the scalars $t^I$ are
associated to their descendants, which we do not consider in the paper.

The linear solutions to the equations of motion turn out to be
\cite{Lee:1998bx}
\begin{eqnarray} h_{\mu\nu}&=&-{6\over 5}\D\, s\, g_{\mu\nu}+{4\over
\D+1}\nabla_{(\mu}\nabla_{\nu)}s\label{hmn}\\
h_{\alpha\beta}&=&2\D\,s\, g_{\alpha\beta}\label{halphabetasol}\\
a_{\mu_1\mu_2\mu_3\mu_4}&=&4\epsilon_{\mu_1\mu_2\mu_3\mu_4\mu_5}\nabla^{\mu_5}b\label{a40}\\
a_{\alpha_1\alpha_2\alpha_3\alpha_4} &=&
-4 \sum_I\epsilon_{\alpha\alpha_1\alpha_2\alpha_3\alpha_4}b^{I}(x)\nabla^{\alpha}Y^{I}(y)
\label{alphaS5} \end{eqnarray}
where $s=\sum s^I Y^I$ and $b=\sum b^I Y^I$.
Using (\ref{halphabeta}) and the solution (\ref{halphabetasol}) one can identify $h_2=10\,\D\,s$.
Setting $t^I=0$ in (\ref{t}), one can then deduce $s=-b$.

\section{Operator product expansion of Wilson loops}
\setcounter{equation}{0}

The Wilson loop operator can be expanded in terms of local operators
when probed from distances much larger than its characteristic size
$a$. For the circular Wilson loop with radius $a$ we can write
\cite{Berenstein:1998ij} \be W(\mathcal{C})=\langle W(\mathcal{C})\rangle
\left(1+\sum_{n} c_{(n)} a^{\D_{(n)}}{\cal O}_{(n)}\right).\ee In
this expression ${\cal O}_{(n)}$ is a local gauge invariant operator
with conformal dimension $\D_{(n)}$, and the sum over $n$ runs over
both the primary operators and their descendants. This operator
product expansion must be invariant under the symmetries preserved
by the Wilson loop. The half-BPS circular loop has
$\theta^I(\tau)=\theta^I=const.$ and therefore preserves a $SO(5)$
subgroup of the original $SO(6)$ R-symmetry group. The operators
appearing in the OPE expansion must therefore contain $SO(5)$
singlets in the $SO(6)\rightarrow SO(5)$ decomposition.  For
example, at level $\D=2$ we can consider the chiral primary operator
${\cal O}^A_2=C_{IJ}^A\mbox{Tr}\Phi^I\Phi^J$, where $C_{IJ}^A$ is a
$SO(6)$ symmetric traceless tensor. Under $SO(6)\rightarrow SO(5)$
it decomposes as ${\bf 20}\rightarrow {\bf 1}+ {\bf 5}+ {\bf 14}$
and therefore, containing a singlet, it will appear in the OPE of
the Wilson operator. A similar analysis can be performed for higher
dimension operators, which in general will contain covariant
derivatives, gauge field-strenghts and the fermions of the
$\mathcal{N}=4$ multiplet. Some of them will get large anomalous
dimension in the strong coupling limit and therefore will decouple.
The generic expansion looks as follows
\begin{eqnarray}\label{OPEs}
\frac{ W(\mathcal{C}) }{\langle W(\mathcal{C}) \rangle }
 &= &  1
+ c_{(2)} \, a^2 \, Y^{(2)}_A(\theta) {\cal N}_2  C^A_{IJ}
\mbox{Tr}\left(\Phi^I \Phi^J\right)+\cr && + c_{(3)} \, a^3 \, Y^{(3)}_A(\theta)
{\cal N}_3 C^A_{IJK} \mbox{Tr}\left(\Phi^I \Phi^J \Phi^K\right) +
c_{(4)} \, a^3  \mbox{Tr}\bigl(\theta^I X^I
F_{+} \bigr)   + \ldots ~\cr &&
\end{eqnarray}
where $Y_A^{(n)}(\theta)$ are spherical harmonics and $\mathcal{N}_n$ are normalization constants.

The coefficients appearing in the OPE expansion can be read off
from the large distance behavior of the two point correlator of
the Wilson loop and the local operators \be {\langle W(\mathcal{C}){\cal
O}^{(n)}(x) \rangle\over \langle
W(\mathcal{C})\rangle}=c_{(n)}{a^{\D_{(n)}}\over L^{2\D_{(n)}}}+\ldots
 \ee where it is
assumed that the loop radius $a$ is much smaller than the distance
$L$ from the point of insertion of the local operator. In this
paper we will focus only on chiral primary operators ${\cal
O}_{\D}^A=C^A_{I_1\cdots I_{\D}}\mbox{Tr}(\Phi^{I_1}\ldots \Phi^{I_{\D}})$ \footnote{We take the  traces of the chiral primaries in the fundamental representation.}. These
belong to short representations of the superconformal algebra, have protected conformal dimensions, and appear at all orders in
the expansion (\ref{OPEs}). 

In the AdS/CFT correspondence the chiral primary operators are dual to supergravity modes:
$\mathcal{O}_{\D}$ corresponds to a scalar of mass $m^2=\D(\D-4)$, which is a combination of the trace of the metric and the RR 4-form over $S^5$, as we reviewed in the previous section.

We now briefly discuss the procedure for computing the correlation function of these operators with a Wilson loop in the strong coupling regime.
The coupling to the
string worldsheet of the supergravity mode dual to $\mathcal{O}_\D$
is given by a vertex operator $V_{\D}$, which can be determined by expanding the
string action to linear order in the fluctuation $h_{\mu\nu}$
\begin{eqnarray}
S&=&{1\over 2\pi\alpha'}\int
d^2\sigma\sqrt{\det\left(G_{\m\n}\partial_{\alpha}x^\m\partial_{\beta}x^\n\right)}\cr
&=&{1\over 2\pi\alpha'}\int
d^2\sigma\sqrt{\det\left(g_{\m\n}\partial_{\alpha}x^\m\partial_{\beta}x^\n\right)}
\left(1+{1\over
2}(g_{\m\n}\partial_{\alpha}x^\m\partial_{\beta}x^\n)^{-1}h_{\m\n}\partial_{\alpha}x^\m\partial_{\beta}x^\n+\ldots\right).\label{sstring}\cr &&
\end{eqnarray}
The fluctuation of the metric $h_{\m\n}$ on $AdS_5$ is given in eq. (\ref{hmn}). We write
the scalar $s^I$ in terms of a source $s^I_0$ located at
the boundary
\bea s^I(\vec{x},z)=\int d^4\vec{x}\,' G_\D(\vec{x}\,';\vec{x},z)s^I_0(\vec{x}\, ')
\label{s1}\eea
where $G_{\D}(\vec{x}\, ';\vec{x},z)$ is the bulk-to-boundary propagator
which describes the propagation of the supergravity mode from the
insertion point $\vec{x}\, '$ of the chiral primary operator to the
point $(\vec{x},z)$ on the string worldsheet \be
G_\D(\vec{x}\, ';\vec{x},z)=c\,\left( z\over
z^2+|\vec{x}-\vec{x}\, '|^2\right)^{\D}.\label{prop0}\ee
The constant $c=\frac{\D+1}{2^{2-\D/2}N\sqrt{\D}}$ is fixed by requiring the unit normalization of the 2-point function
\cite{Berenstein:1998ij}.
Since we are probing the Wilson loop from
a distance $L$ much larger than its radius $a$ we can approximate
\bea G_\D(\vec{x}\, ';\vec{x},z) \simeq c\,{z^{\D}\over L^{2\D}}\, , ~~~~~~~~ \p_z
s^I\simeq\frac{\D}{z}s^I\, , ~~~~~~~~ \p^2_z s^I\simeq \frac{\D
(\D-1)}{z^2} s^I.\label{s2}\eea The relevant Christoffel symbols are readily
computed to be \bea \Gamma^z_{\m\n}= z g_{\m\n}
-\frac{2}{z}\delta^z_\m \delta^z_\n \label{christoffel}\eea so that
one finally has \bea h_{\m\n}^I\simeq -2\D ~ g_{\m\n} s^I+\frac{4
\D}{z^2} \delta^z_\m \delta^z_\n s^I. \label{hfinal}\eea
Inserting this result into (\ref{sstring}), the coupling to the worldsheet is found to be
\cite{Berenstein:1998ij} \be
{1\over 2\pi\alpha'}\int d{\cal A}\,(-2\D\, s){z^2\over a^2}
\equiv {1\over 2\pi\alpha'}\int d{\cal A}\, V_\D\, s.\ee
In this expression $d{\cal A}$ is
the area element of the classical string.
The correlation function is obtained from functionally differentiating
the previous formula with respect to the source $s_0$
\bea
{\langle W(\mathcal{C})
{\cal O}_{\D}(\vec{x}_0)\rangle\over \langle
W(\mathcal{C})\rangle}&=&-Y^I(\theta)\frac{\delta}{\delta s_0(\vec{x}_0)}\frac{1}{2\pi\alpha'}
\int d\mathcal{A} \, d^4 x'\,  V_\D \,G_\D(\vec{x}\,';\vec{x},z)s^I_0(\vec{x}\, ')\cr
&=&-Y^I(\theta){1\over 2\pi \alpha'}\int d{\cal A}\,\, V^{\D}
G_{\D}(\vec{x}_0;\vec{x},z).\eea
One obtains in the approximations of eq. (\ref{s2})
\be {\langle W(\mathcal{C})
{\cal O}_{\D}(\vec{x}_0)\rangle\over \langle
W(\mathcal{C})\rangle}=2^{\D/2-1}{\sqrt{\D\lambda}\over N} {a^{\D}\over
L^{2\D}}.\label{cstringa}\ee

We now move on to studying the operator product expansion of Wilson
loops in higher dimensional representations. We analyze the rank $k$
symmetric representation first. In the bulk this is described by a
D$3_k$ brane.

\section{Brane computation}
\setcounter{equation}{0}

\subsection{The D3 brane}

We consider a small circular Wilson loop of radius $a$ placed on the
boundary of $AdS_5$. The metric on $AdS_5$ can be written in polar
coordinates as
\bea
ds^2_{AdS}=\frac{1}{z^2}\left(dz^2+dr_1^2+r^2_1 d\psi^2 + dr_2^2 + r^2_2 d\phi^2\right).
\label{AdSpoincare}\eea
The position of the loop is defined by $r_1=a$ and $z=r_2=0$.
We take a D3 brane which pinches off on this circle as $z\rightarrow 0$ and preserves a
$SO(2,1)\times SO(3)\times SO(5)$ isometry \cite{Drukker:2005kx}.

The bulk action includes a DBI part and a Wess-Zumino term, which
captures the coupling of the background Ramond-Ramond field to the
brane \bea S_{D3}=T_{D3}\int\sqrt{\det(\gamma + 2\pi\alpha'
F)}-T_{D3}\int P[C_{(4)}]\label{S00}\eea where
$T_{D3}=\frac{N}{2\pi^2}$ is the tension of the brane, $\gamma$ is
the induced metric, $F$ the electromagnetic field strenght, and
$P[C_{(4)}]$ is the pull-back of the 4-form \bea C_{(4)}=\frac{r_1
r_2}{z^4}dr_1\wedge d\psi \wedge dr_2\wedge d\phi\label{C4}\eea
to the brane worldvolume.

We review the brane solution found in \cite{Drukker:2005kx}. It
turns out to be more convenient to use a new set of coordinates
obtained by transforming $\{z,\, r_1,\, r_2\}$ into \bea
z=\frac{a\sin\eta}{\cosh\rho-\sinh\rho\cos\theta}\, , ~~
r_1=\frac{a\cos\eta}{\cosh\rho-\sinh\rho\cos\theta}\, , ~~
r_2=\frac{a\sinh\rho\sin\theta}{\cosh\rho-\sinh\rho\cos\theta}.
\label{z}\eea In this coordinate system the metric on $AdS_5$ reads
\bea ds^2_{AdS}=\frac{1}{\sin^2\eta}\left(d\eta^2+\cos^2\eta ~
d\psi^2 +d\rho^2 +\sinh^2\rho ~(d\ta^2+\sin^2\ta ~
d\f^2)\right)\label{metricAdS}\eea where $\rho\in [0,\infty)$,
$\theta\in [0,\pi]$, and $\eta\in [0, \pi/2]$. The Wilson loop is
located at $\eta=\rho=0$. One can pick a static gauge in which the
worldvolume coordinates of the brane are identified with $\{\psi\, ,
\rho\, ,\ta\, , \f \}$ and the brane sits at a fixed point of the
$S^5$ determined by the constant unit vector $\theta^I\in
\mathbb{R}^6$. The remaining coordinate can be seen as a scalar
field, $\eta=\eta(\rho)$. Because of the symmetries of the problem
the electromagnetic field has only one component,
$F_{\psi\rho}(\rho)$. In this coordinates the DBI action in
(\ref{S00}) reads \bea S_{DBI}=2N\int d\rho d\theta \,
\frac{\sin\theta\sinh^2\rho}{\sin^4\eta}
\sqrt{\cos^2\eta(1+\eta'^2)+(2\pi\alpha')^2\sin^4\eta
F_{\psi\rho}^2}\label{DBI00}\eea while the Wess-Zumino term is \bea
S_{WZ}=-2N\int d\rho d\theta \,
\frac{\cos\eta\sin\theta\sinh^2\rho}{\sin^4\eta} \left(\cos\eta
+\eta'\sin\eta\,\frac{\sinh\rho-\cosh\rho\cos\theta}
{\cosh\rho-\sinh\rho\cos\theta}\right). \label{WZ00}\eea The
solution to the equations of motion reads \cite{Drukker:2005kx} \bea
\sin\eta=\frac{1}{\K}\sinh\rho\, , ~~~~~~~~ F_{\psi\rho}=\frac{i k
\lambda}{8\pi N \sinh^2\rho}\, ,
~~~~~~~~\K=\frac{k\sqrt{\lambda}}{4N}. \label{ansatzD3}\eea

The bulk action has to be complemented with boundary terms for the
worldvolume scalar $\eta$ and for the electric field $F_{\psi\rho}$
\cite{Drukker:2005kx}. These terms do not change the solution but
alter the final value of the on-shell action which reads \bea
S_{D3}=S_{DBI}+S_{WZ}+S_{\mbox{\tiny{bdy}}}=-2N(\K\sqrt{1+\K^2}+\sinh^{-1}\K).\eea
The expectation value of a Wilson loop in the rank $k$ symmetric
\footnote{In the limit of $N\rightarrow\infty$ and
$\lambda\rightarrow\infty$ the symmetric representation coincides
with the multiply wound Wilson loop
\cite{Okuyama:2006jc}\cite{Hartnoll:2006is}.} representation is then
\bea \LA W_{S_k}\RA =
\exp\left(2N(\K\sqrt{1+\K^2}+\sinh^{-1}\K)\right).\eea For small
$\K$ this expression reproduces the result of $k$ fundamental
strings \bea \LA W_{S_k}\RA \simeq e^{k\sqrt{\lambda}}.\eea

\subsubsection{Coupling to chiral primaries}

The linearized coupling of the scalar $s^I$ to the brane
worldvolume can be found by expanding the induced metric on the
brane around the $AdS_5\times S^5$ background $g_{mn}$ and keeping the
first order term in the
fluctuation $h_{mn}$. Since
the brane lies completely in $AdS_5$ we can write \bea
S_{DBI}& =& T_{D3}\int ~ d^4 \sigma
\sqrt{\det\left(G_{\m\n}\p_{a}X^\m\p_{b}X^\n+2\pi\alpha' F_{a
b}\right)}\cr &=& T_{D3}\int ~ d^4 \sigma
\sqrt{\det\left(g_{\m\n}\p_{a}X^\m\p_{b}X^\n+2\pi\alpha' F_{a
b}\right)}~\cdot \cr && \hskip1.5cm \cdot
\left(1+\frac{1}{2}\left(g_{\m\n}\p_{a}X^\m\p_{b}X^\n+2\pi\alpha'
F_{a b}\right)^{-1}h_{\rho\sigma}\p_{a}X^\rho
\p_{b}X^\sigma+\ldots\right).\label{DBI1}\eea Here $a, \, b$ are the brane
worldvolume indices.

The coupling
of $s^I$ to the 4-form in the Wess-Zumino term is obtained by replacing $C_{(4)}\rightarrow
C_{(4)}+a_{(4)}$ where, using eq. (\ref{a40})
and the approximation (\ref{s2}), the fluctuation $a_{(4)}$ is \bea
a^I_{\m_1\ldots\m_4}\simeq-4\epsilon_{\m_1\ldots\m_4
z}\p^{\,z}s^I\simeq-4\D\, z \, \epsilon_{\m_1\ldots\m_4 z}
s^I\label{a2}\eea so that \bea S_{WZ}^{(1)}=-T_{D3} \int
P[a_{(4)}]=4\, T_{D3}\, \D\, \int P[C_{(4)}]s\label{WZ2}\eea
where $s=\sum s^I Y^I$.

We use now the explicit solution to the equations of motion
(\ref{ansatzD3}) to evaluate the on-shell value of the fluctuations
(\ref{DBI1}) and (\ref{WZ2}). 
The first order in the fluctuation in (\ref{DBI1}) turns out to be \bea
S_{DBI}^{(1)}=4 N \D \K^2 \int d\rho d\ta\,
\frac{\sin\ta}{\sinh^2\rho}\left(-1-2\K^2+\frac{1-\sinh^2\rho
(\K^{-2}-\sin^2\ta)}{(\cosh\rho-\sinh\rho\cos\ta)^2}\right)s.\label{DBI2}\eea
Similarly, the
Wess-Zumino term reads \bea S_{WZ}^{(1)}= 8 N \D \K^4 \int d\rho d\ta\,
\frac{\sin\ta}{\sinh^2\rho}\left(1+\frac{1}{\K^2}\frac{\sinh^3\rho
-\sinh\rho\cosh^2\rho}{\cosh\rho-\sinh\rho\cos\ta}\cos\theta\right)s.\label{WZ3}\eea
The final result for the action is then
\bea S^{(1)}_{D3}=S^{(1)}_{DBI}+S^{(1)}_{WZ}=-4 N \D
\int_0^{\sinh^{-1}\K} d\rho \int_0^{\pi} d\ta \,
\frac{\sin\ta}{(\cosh\rho-\sinh\rho\cos\ta)^2}s.\label{Sfinal}\eea

\subsubsection{The correlation function}

The prescription for computing the correlation function between the
Wilson loop and the chiral primary operator is to functionally
differentiate the action (\ref{Sfinal}) with respect to
the source $s^I_0$ (see eq. (\ref{s1})) \bea \frac{\LA
W(\mathcal{C}) \mathcal{O}_\D(L)\RA}{\LA W(\mathcal{C})\RA}= -
\frac{\delta S_{D3}^{(1)}}{\delta s_0}\Big|_{s_0=0}.\label{OPE0}\eea

We approximate the bulk-to-boundary propagator with
$c\frac{z^\D}{L^{2\D}}$ and use for $z$ the expression (\ref{z}).
This yields \bea \frac{\LA W(\mathcal{C}) \mathcal{O}_\D(L)\RA}{\LA
W(\mathcal{C})\RA}&\simeq
&\frac{a^\D}{L^{2\D}}\,\frac{4N\D}{\K^\D}\, c
\,\int_0^{\sinh^{-1}\K} d\rho \sinh^\D \rho \int_0^{\pi} d\ta
\frac{\sin\ta}{(\cosh\rho-\sinh\rho\cos\ta)^{2+\D}}.\cr
&&\label{OPE}\eea  We are
neglecting terms of higher order in $\frac{a}{L^2}$.

After performing the two integrals, the final result for the
coefficients of the operator product expansion turns out to be remarkably simple \bea
c_{S_{k},\D}=\frac{2^{\D/2+1}}{\sqrt{\D}}\sinh(\D\sinh^{-1}\K).\label{c}\eea
Interestingly enough, this can be expressed in terms of Chebyshev
polynomials with imaginary argument \bea
c_{S_k,\D}=\frac{(-1)^{\D/2}
2^{\D/2+1}}{\sqrt{\D}}\cdot\left\{\begin{array}{lc} -iV_{\D}(i\K) & \mbox{   for } \D \mbox{ even}\\
T_{\D}(i\K) & \mbox{   for } \D \mbox{
odd}\end{array}\right.\label{ccheb}\eea where we have used the identities
$T_n(x)=\cos(n\cos^{-1}x)$ and $V_{n}(x)=\sin(n\cos^{-1}x)$.

The string limit is recovered when $\K\rightarrow 0$. In this regime
the $S^2$ in the brane worldvolume shrinks to zero size and the D3
reduces effectively to a fundamental string with $AdS_2$
worldsheet. The coefficients (\ref{c}) become \bea c_{S_k,\D}\simeq
2^{\D/2+1} \sqrt{\D} \, \K = 2^{\D/2-1}  \,
\frac{\sqrt{\D\lambda}}{N}\, k\label{cstring}\eea in perfect agreement
with the result (\ref{cstringa}) found originally in \cite{Berenstein:1998ij}.

\subsection{The D5 brane}

Circular Wilson loops in the rank $k$ antisymmetric representation
of the gauge group have a bulk description in terms of D5 branes
with $AdS_2 \times S^4$ worldvolume and $k$ units of fundamental
string charge dissolved in them \cite{Yamaguchi:2006tq}
\cite{Gomis:2006sb}. The $D5$ description of these Wilson loops is
valid in the large $N$, large $\lambda$ limit with $k/N$ fixed.
Before moving on to compute the coupling of these branes to the
scalars $s^I$ dual to chiral primaries, we briefly review the D5
solution \cite{Yamaguchi:2006tq} to set up the notation and our
conventions. It is convenient to take the $AdS_5 \times S^5$ metric
as \bea
ds^2 &= & \mbox{cosh}^2u (d \zeta^2 + \mbox{sinh}^2 \zeta d \psi^2) + du^2 + \mbox{sinh}^2 u (d \vartheta^2 + \mbox{sin}^2               \vartheta d \phi^2)  + \nonumber \\
        && + \, \, d \theta^2 + \mbox{sin}^2 \theta d \Omega_4^2 \, ,
\label{Ads2S2} \eea where we have written the $AdS_5$ factor as an
$AdS_2 \times S^2$ fibration. These coordinates are related to the
usual Poincare patch by \bea r_1 &=& {a \cosh u \sinh \zeta \over
\cosh u \cosh \zeta -\cos \vartheta \sinh u}\, ,~~~~~ \quad
r_2 = {a \sinh u \sin \vartheta \over \cosh u \cosh \zeta -\cos \vartheta \sinh u} \nonumber \\
z &=& {a \over \cosh u \cosh \zeta -\cos \vartheta \sinh u}
\label{Ads2S2coord} \eea where, as before, $a$ denotes the radius of the Wilson loop. 
In these coordinates, the Wilson loop is at
$\zeta \rightarrow \infty$, $u=0$ and it is parametrized by $\psi$.
The selfdual 4-form potential can be taken to be \bea C_{(4)} = 4
\, \left( {u \over 8} - {1 \over 32} \sinh 4u \right) \, dH_2 \wedge
d \Omega_2 - \left({3 \over 2} \, \theta - \sin 2 \theta + {1 \over
8} \sin 4 \theta \right) \, d \Omega_4 \, , \label{D5C4} \eea where
$dH_2$ denotes the volume element of the $AdS_2$ part of the metric.

Since we want to construct a $D5$ brane with $AdS_2 \times S^4$ worldvolume, it is natural to take a static
gauge in which $\zeta,\psi$ and the coordinates of the $S^4 \subset S^5$ are the worldvolume coordinates.
Furthermore we can take the following ansatz which preserves the $SO(2,1) \times SO(3) \times SO(5)$ symmetry
of the Wilson loop
\bea
u=0\, ,~~~~~~~~~ \theta = const.
\eea
and only the $F_{\psi \zeta}$ component of the worldvolume gauge field is turned on. With this ansatz the DBI
and Wess-Zumino parts of the D5 action reduce to
\bea
S_{DBI} &=& \, T_{D5} \int d^6 \sigma \sqrt{\mbox{det} (\gamma_{ab} + 2 \pi \alpha' F_{ab})} \nonumber \\
        &=& \, {2 N \over 3 \pi} \sqrt{\lambda} \int \, d \zeta \sinh\zeta \sin^4\theta\sqrt{ 1 + {4 \pi^2 \over \lambda}
{F_{\psi \zeta}^2 \over \sinh^2 \zeta}} \, ,
\label{D5DBI}
\eea
\bea
S_{WZ} &=& \, - 2 \pi \alpha' i \, T_{D5} \int F \wedge P[C_{(4)}] \nonumber \\
       &=& \, {4 i N \over 3} \int d \zeta \, F_{\psi \zeta}\,  \left( {3 \over 2} \, \theta - \sin 2 \theta + {1 \over 8}
\sin 4 \theta \right) \label{D5WZ} \eea where we have used
$T_{D5} = N \sqrt{\lambda} / 8 \pi^4$ and $\mbox{vol}(\Omega_4) = 8
\pi^2 / 3$. The equation of motion for the electric field states
that the conjugate momentum is a constant equal to the number of
fundamental string charge $k$ dissolved in the D5 brane \bea \Pi
\equiv {-i \over 2 \pi} \frac{\delta \cal{L}}{\delta F_{\psi \zeta}}
= {2 N \over 3 \pi} {E \sin^4 \theta \over \sqrt{1-E^2}} + {2 N
\over 3 \pi} \left( {3 \over 2} \, \theta - \sin 2 \theta + {1 \over
8} \sin 4 \theta \right) = \, k \label{D5PI} \eea where for
convenience we have defined $E = {- 2 \pi i \over \sqrt{\lambda}}
{F_{\psi \zeta} \over \sinh \zeta}$. This equation allows to
determine the angle $\theta$ at which the D5 sits as a function of
$k$ \bea \theta_k - \sin \theta_k \, \cos \theta_k = \, \pi\, {k \over
N} \label{thetak} \eea while the electric field is given by $E =
\cos \theta_k$. One can check that with this ansatz the equation of
motion for $u$ is also satisfied. Adding the appropriate boundary
terms for the electric field and the worldvolume scalars (see
\cite{Yamaguchi:2006tq}\cite{Hartnoll:2006ib} for details) the
on-shell action for the D5 brane becomes \bea S_{D5} = S_{DBI} +
S_{WZ} + S_{\mbox{\tiny{bdy}}} = \, - {2 N \over 3 \pi}
\sqrt{\lambda} \sin^3 \theta_k  \eea so the expectation value of
the Wilson loop in the rank $k$ antisymmetric representation is
given by \bea \langle W_{A_k} \rangle = \, \mbox{exp} \left({2 N
\over 3 \pi} \sqrt{\lambda} \sin^3 \theta_k \right) \, .
\label{antiwilson} \eea As previously noted in the literature, this
result is consistent with the duality between the rank $k$ and rank
$N-k$ antisymmetric representations: indeed, it can be seen from eq.
(\ref{thetak}) that under $k \rightarrow N-k$ the angle $\theta_k$
goes into $\pi - \theta_k$.  It can also be checked that in the
limit $k/N \rightarrow 0$, in which the $S^4$ factor shrink to zero
size,  (\ref{antiwilson}) coincides with the action of $k$
fundamental strings, as for small $k/N$ eq. (\ref{thetak}) gives
$\theta_k^3 \sim 3 \pi k / 2N$, so that $\langle W_{A_k} \rangle
\simeq \mbox{exp} \, k \sqrt{\lambda}$.

\subsubsection{Coupling to chiral primaries}
The coupling of the KK scalars $s^I$ to the D5 worldvolume can be obtained along
the same lines of the D3 calculation of the previous section. However, besides the
fluctuation of the $AdS_5$ part of the metric $h_{\mu\nu}$, we also need
the fluctuation of the metric in the $S^5$ direction $h_{\alpha\beta}$ as well
as the fluctuation of the 4-form $a_{(4)}$ along the $S^4$. The explicit
expressions can be found in section 2.
In particular, in this coordinates the 4-form over the $S^5$ is
\bea
a_{\sigma_1\sigma_2\sigma_3\sigma_4} = 4 \sin^4 \theta \,
\mu(\Omega_4) \sum s^{I} \, \partial_{\theta}
Y^{I} \label{D5fluct} \eea where $\sigma_1, \ldots, \sigma_4$ are the coordinates on the $S^4$ and
$\mu(\Omega_4)= \sin^3 \sigma_1 \sin^2 \sigma_2 \sin \sigma_3$ is
the corresponding measure. Differently from the $D3$ compuation, in
this case the $S^5$ spherical harmonics $Y^{I}$ play an
active role in the computation since the $D5$ brane extends into the
5-sphere. The explicit form of the harmonics is given in the
appendix.

The variation of the DBI part of the action to first order in the
fluctuations $h_{\mu\nu}$ and $h_{\alpha\beta}$ reads \bea
S_{DBI}^{(1)} = \, { T_{D5} \over 2} \int \sqrt{\mbox{det}
(\gamma_{ab} + 2 \pi \alpha' F_{ab})} \, \big( \gamma_{ab} + 2 \pi
\alpha' F_{ab} \big )^{-1} \big (h_{\mu\nu} \partial_a X^{\mu}
\partial_b X^{\nu} + h_{\alpha\beta} \partial_a X^{\alpha}
\partial_b X^{\beta} \big ). \nonumber \\
\eea Using the explicit solution reviewed in the previous section,
it is easy to compute the matrix $\gamma_{ab}+ 2 \pi \alpha'
F_{ab}$. Plugging in the explicit expressions for the fluctuations
and using the fact that on the D5 solution we have $z= a / \cosh
\zeta$ (this follows from the change of coordinate
(\ref{Ads2S2coord}) after setting $u=0$), we get after some
computations \bea S_{DBI}^{(1)} = \pi T_{D5} \int d \zeta d\sigma_1
\ldots d\sigma_4 \mu (\Omega_4) \sinh \zeta \sin^5 \theta_k \left (
-{4 \Delta \over \cosh^2 \zeta \sin^2 \theta_k} + 8 \Delta \right )
s^{I} Y^{I}. \label{D5DBI1} \eea Performing the
integration over the $S^4$, only the $SO(5)$ invariant spherical
harmonics are selected, namely the harmonics which depends on
$\theta_k$ only, and we get \bea S_{DBI}^{(1)} = {N \sqrt{\lambda}
\over 3 \pi} \int d \zeta \sinh \zeta \sin^5 \theta_k \left ( -{4
\Delta \over \cosh^2 \zeta \sin^2 \theta_k} + 8 \Delta \right )
s^{\Delta} Y^{\Delta, 0} (\theta_k) \label{D5DBI1_f} \eea where
the suffix on the harmonic indicates that all the quantum numbers
except one were set to zero by the integration over the 4-sphere. As
reviewed in the appendix, these $S^4$ invariant harmonics can be
explicitely written as \bea Y^{\Delta, 0} (\theta_k) = \, {\cal
N}_{\Delta} \,  C_{\Delta}^{(2)} \, (\cos \theta_k)
\label{harmonicgege} \eea where $C_{\Delta}^{(2)} \, (\cos
\theta_k)$ are Gegenbauer polynomials, and ${\cal N}_{\Delta}$ is a
normalization constant necessary to have orthonormality.

The linear coupling coming from the Wess-Zumino part of the action (\ref{D5WZ}) can be obtained using the
expression for the 4-form fluctuation in eq. (\ref{D5fluct}), and after integrating over the $S^4$ as above, we get
\bea
S_{WZ}^{(1)} = \, {8 N \sqrt{\lambda} \over 3 \pi} \int d\zeta \sinh \zeta \sin^4 \theta_k \cos \theta_k
s^{\Delta} \, \partial_{\theta_k} Y^{\Delta , 0} (\theta_k).
\label{D5WZ1_f}
\eea

\subsubsection{The correlation function}
The correlator between the rank $k$ antisymmetric Wilson loop and
chiral primary operator ${\cal O}_{\Delta} (L)$ can now be computed
plugging (\ref{s1}) into (\ref{D5DBI1}) and (\ref{D5WZ1_f}) and
differentiating with respect to the source $s_0^{\Delta}$. As
before, the bulk-to-boundary propagator can be approximated by $c\, z^\D / L^{2
\Delta}$. Recalling that on the D5 solution $z= a / \cosh \zeta$,
the $\zeta$-integrals can be readily computed and we get \bea
\frac{\langle W_{A_k} \, {\cal O}_{\Delta} (L) \rangle}{\langle
W_{A_k} \rangle} &=& {a^{\Delta} \over L^{2 \Delta}} \bigg [
{2^{\Delta / 2} \over 3 \pi} \sqrt{\Delta \lambda} \sin^3 \theta_k
Y^{\Delta, 0} (\theta_k) - \cr && - {2^{\Delta / 2+1} \sqrt{\lambda}
(\Delta + 1) \over 3 \pi \sqrt{\Delta} (\Delta -1)} \sin^5 \theta_k
\left(\Delta Y^{\Delta, 0} (\theta_k) + {\cos \theta_k \over \sin
\theta_k} \partial_{\theta_k} Y^{\Delta, 0} (\theta_k) \right) \, \bigg
].\cr && \label{corrD5_1} \eea Using the formula for the derivatives of
Gegenbauer polynomials eq. (\ref{gegenderivapp}), we obtain \bea
\Delta Y^{\Delta, 0} (\theta_k) + {\cos \theta_k
\over \sin \theta_k}
\partial_{\theta_k} Y^{\Delta, 0} (\theta_k) =
 {{\cal N}_{\Delta} \over \sin^2 \theta_k} \big ( \Delta C_{\Delta}^{(2)} (\cos \theta_k) -
(\Delta +3) \cos \theta_k \, C_{\Delta-1}^{(2)} (\cos \theta_k) \big
). \cr &&\label{gegederiv} \eea The correlation function
(\ref{corrD5_1}) can then be written as \bea \! \! \! \frac{\langle
W_{A_k} \, {\cal O}_{\Delta} (L) \rangle}{\langle W_{A_k} \rangle}
&=& {a^{\Delta} \over L^{2 \Delta}} Y^{\Delta, 0} (0)  \bigg
[{2^{\Delta / 2} \over 3 \pi} \sqrt{\Delta \lambda} \sin^3 \theta_k
\cdot \cr &&  \cdot \, {6 (\Delta -2) !  \over (\Delta + 2)!} 
\left(2 (\Delta +1) \cos \theta_k \, C_{\Delta-1}^{(2)} (\cos \theta_k) -
\Delta C_{\Delta}^{(2)} (\cos \theta_k) \right) \bigg ]
\label{corrD5_2} \eea where we have factorized out the spherical
harmonic evaluated at $\theta=0$, $Y^{\Delta, 0} (0) = {\cal
N}_{\Delta} {(\Delta +3)! \over 6 \Delta !}$ \footnote{The OPE
coefficient does not include a factor coming from the spherical
harmonic evaluated at the unit 6-vector $\theta^I$ appearing in eq.
(\ref{wilson0}). After a rotation, this vector can always be set to
$\theta^I = (1,0,\ldots,0)$ which corresponds to the north pole of
$S^5$, $\ie$ $\theta=0$.}. The OPE coefficient $c_{A_k, \Delta}$ we
aim to compute is the expression in square brackets. Using the
recurrence relation eq. (\ref{gegenrecurr}), we find that this
expression can be written in the compact form \bea c_{A_k, \Delta} =
\, {2^{\Delta / 2} \over 3 \pi} \sqrt{\Delta \lambda} \sin^3
\theta_k \,
 {6 (\Delta -2) !  \over (\Delta + 1)!} \,  C_{\Delta-2}^{(2)} (\cos \theta_k).
\label{corrD5_fin} \eea This is our final result for the correlation
function of rank $k$ antisymmetric Wilson loops and chiral
primaries. In the next section, we will see that this result exactly
matches the one obtained from the normal matrix model. As a check,
one can verify that this expression reduces to the string result of
\cite{Berenstein:1998ij} in the limit $k/N \rightarrow 0$, by using
$\theta_k^3 \sim 3 \pi k / 2N$ and
eq. (\ref{C1}) from the appendix.

\section{The correlation functions from the normal matrix model}
\setcounter{equation}{0}

It is well-known that the expectation value of a circular Wilson
loop in the fundamental representation of $SU(N)$ can 
be computed from a quadratic Hermitian matrix model
\cite{Erickson:2000af}\cite{Drukker:2000rr} \bea \LA
W_{\square}\RA=\frac{1}{\mathcal{Z}_H}\int
[dM]\exp\left(-\frac{2N}{\lambda}\mbox{Tr}
M^2\right)\frac{1}{N} \mbox{Tr}_{\square}\, 
e^M.\label{mm0}\eea This matrix model is conjectured to capture the
physics of the Wilson loop exactly, up to instanton corrections
\cite{Bianchi:2002gz}, to all orders of $1/N$ and $\lambda$. The conjecture
extends to higher rank Wilson loops as well. The
result for the multiply wound Wilson loop has been obtained in
\cite{Drukker:2005kx}, whereas \cite{Hartnoll:2006is} and
\cite{Yamaguchi:2006tq}\cite{Hartnoll:2006is} contain the
computations for, respectively, the symmetric and antisymmetric
representations.

When computing the correlation function between a Wilson loop and a
chiral primary operator one can 
substitute the Hermitian model (\ref{mm0}) with a complex one by introducing 
a second matrix $M_{\mbox{\tiny{Im}}}$ and defining $z=M+iM_{\mbox{\tiny{Im}}}$.
In \cite{Okuyama:2006jc} it was shown that, for certain 
representations of the Wilson loop (the multi-winding and the antisymmetric), 
the complex matrix model is equivalent to a normal
matrix model, which is a complex model where the matrix is constrained to commute with its
conjugate. 
In the normal matrix model the expression for the Wilson
loop reads \bea \LA
W_{\mathcal{R}}\RA=\frac{1}{\mathcal{Z}_N}\int_{[z,\bar z]=0} [d^2
z]\exp\left(-\frac{2N}{\lambda}\mbox{Tr} z\bar
z\right)\frac{1}{\mbox{dim} \mathcal{R}} \mbox{Tr}_{\mathcal{R}}
e^{\frac{1}{\sqrt{2}}(z+\bar z)-\frac{\lambda}{8N}}.\label{mm2}\eea 

For large $N$, the
eigenvalues of this model are distributed in incompressible droplets in the
complex plane. This leads to interpreting the complex plane as the phase space of free fermions, in analogy 
with 
the matrix quantum mechanics describing
chiral primary operators \cite{Berenstein:2004kk}\cite{Lin:2004nb}. 
For example,  the
Wilson loop in the fundamental representation has an eigenvalue
distribution given by a circular droplet with constant density \footnote{Projecting (\ref{rhoz}) into the real axis one recovers the Wigner semi-circle distribution.}\bea \rho(z)=
\left\{ 
\begin{matrix} \frac{2}{\pi\lambda} & ~~~~~
|z|<\sqrt{\frac{\lambda}{2}} \cr 0 & ~~~~~ |z|>\sqrt{\frac{\lambda}{2}}
 \end{matrix}
\right.
\label{rhoz}\eea

In \cite{Okuyama:2006jc} it was also shown that the correlation function
between a Wilson loop in the fundamental representation and a chiral
primary operator is given by \footnote{The factor $2^{\Delta/2}$ instead of the $2^{-\Delta/2}$ of \cite{Okuyama:2006jc} is 
set in order to have normalizations consistent with \cite{Berenstein:1998ij}.}
\bea \LA W_{\square}\mathcal{O}_\D \RA=
\frac{2^{\D/2}}{\mathcal{Z}_N}\int_{[z,\bar z]=0} [d^2 z]
\exp\left(-\mbox{Tr}\,z\bar
z\right)\frac{1}{N}\mbox{Tr}_{\square}\,e^{\frac{1}{2}\sqrt{\frac{\lambda}{N}}(z+\bar
z)-\frac{\lambda}{8N}}\frac{1}{\sqrt{\D N^{\D}}}\mbox{Tr}
\,z^\D.\label{wo}\eea
We now use the normal matrix model to check our results for the coefficients of the operator product expansion of higher rank Wilson loops.

\subsection{The symmetric case}

We start by reproducing the result (\ref{c}) for $c_{S_k,\D}$ using the
normal matrix model. According to the holographic
dictionary put forward in \cite{Gomis:2006sb}, we are interested in
the correlator between a Wilson loop in the rank $k$ symmetric
representation and the chiral primary operator
$\mathcal{O}_{\D}=\frac{1}{\sqrt{\D N^{\D}}}\mbox{Tr}Z^\D$. In the
limit of large $N$ and large $\lambda$ the symmetric representation
Wilson loop $W_{S_{k}}$ effectively coincides with the multiply
wound fundamental loop $W^{(k)}_{\square}$, as was shown in
\cite{Okuyama:2006jc} and \cite{Hartnoll:2006is}. Therefore we limit
ourselves to the simpler case of computing $\LA W^{(k)}_{\square}
\mathcal{O}_\D\RA$, where $k$ is the winding number and corresponds
in the brane probe picture to the number of fundamental strings
dissolved in the brane.

We start from eq. (4.7) of \cite{Okuyama:2006jc}, where we replace
everywhere $\lambda\rightarrow k^2\lambda$ \bea \LA
W^{(k)}_{\square} \mathcal{O}_\D\RA =
\frac{2^{\Delta/2+1} e^{k^2\lambda/8N}}{k\sqrt{\D\lambda}}\oint\frac{dw}{2\pi
i}w^\D e^{k\sqrt{\lambda}\,
w/2}\left(1+\frac{k\sqrt{\lambda}}{2Nw}\right)^N
\left[\left(1+\frac{k\sqrt{\lambda}}{2Nw}\right)^\D-1\right]\label{mm}\eea
The large winding limit consists in taking $N\rightarrow\infty$
while keeping $\K\equiv\frac{k\sqrt{\lambda}}{4N}$ fixed. In this
limit the integral can be evaluated around the saddle point of the
terms proportional to $N$ and $k$ \bea
\p_w\left(\frac{k\sqrt{\lambda}}{2}\, w + N \log
\left(1+\frac{k\sqrt{\lambda}}{2Nw}\right)\right)=0\label{saddle0}\eea
which yields \bea w_{\star}=\sqrt{1+\K^2}-\K\label{w}.\eea Inserting $w_*$
in (\ref{mm}) and using  \bea \sqrt{1+\K^2}+\K=\exp(\sinh^{-1}\K),
~~~~~~~~ \sqrt{1+\K^2}-\K=\exp(-\sinh^{-1}\K)\label{expsinh}\eea it
is easy to see that \bea \LA W^{(k)}_{\square} \mathcal{O}_\D\RA =
\frac{2^{\Delta/2}}{2N\K\sqrt{\D}}\,
2\sinh(\D\sinh^{-1}\K)e^{2N(\K\sqrt{1+\K^2}+\sinh^{-1}\K)}.\label{OPE2}\eea
To get a properly normalized expression one still needs to divide
 (\ref{OPE2}) by  
\bea\LA W^{(k)}_{\square}\RA=
\frac{1}{2N\K}e^{2N(\K\sqrt{1+\K^2}+\sinh^{-1}\K)}.\eea The final
result coincides with eq. (\ref{c}), which we obtained from the
brane picture.

\subsection{The antisymmetric case}

To compute the OPE coefficients of Wilson loops in the rank $k$ antisymmetric representation, we
have to evaluate the following correlator in the normal matrix model \cite{Okuyama:2006jc}
\bea
\LA W_{A_k} \, {\cal O}_{\Delta} \RA = \frac{2^{\Delta /2} \, e^{k \lambda / 8 N}}{\mathcal{Z}_N N^{\Delta /2} \sqrt{\Delta}}
\int_{[z,\bar z]=0} [d^2 z] e^{-\mbox{\tiny Tr} (z \bar z)}
\frac{1}{\mbox{dim} A_k} \mbox{Tr}_{A_k} e^{\frac{1}{2} \sqrt{\frac{\lambda}{N}} (z+\bar z)}\, \mbox{Tr} z^{\Delta}.
\label{nmmantisymm}
\eea
This matrix integral can actually be solved exactly, as was shown in  \cite{Okuyama:2006jc}, and similarly to the case
of the fundamental representation, it can be written as a $k$-dimensional contour integral. However, it does not seem to
be easy to take the large $N$ and large $k$ limit with $k/N$ fixed from such an expression. Here we follow a different
approach to get the above correlator in this limit.  First, as in \cite{Hartnoll:2006is}, we find it convenient to rewrite
the trace in the antisymmetric representation using the corresponding generating function
\bea
\mbox{Tr}_{A_k} e^{\frac{1}{2} \sqrt{\frac{\lambda}{N}} (z+\bar z)} =
\oint \frac{dt}{2 \pi i} t^{k-1} \exp{\left[ \mbox{Tr} \log \left(1+ \frac{1}{t} \, e^{\frac{1}{2} \sqrt{\frac{\lambda}{N}} (z+\bar z)}\right)\right]}.
\label{antigen}
\eea
Since we expect the correlator to be real, it is also convenient to replace $\mbox{Tr} z^{\Delta} \rightarrow \frac{1}{2} (\mbox{Tr} z^{\Delta} + \mbox{Tr} \bar z^{\Delta})$. The idea is then to view the insertion of the chiral primary $\mbox{Tr} z^{\Delta}$ in (\ref{nmmantisymm}) as a small perturbation of the gaussian potential, by writing
\bea
&& \int_{[z,\bar z]=0} [d^2 z] e^{-\mbox{\tiny Tr} (z \bar z)} \mbox{Tr} z^{\Delta}\, e^{\mbox{\tiny Tr} \log (1+ \frac{1}{t} \, e^{\frac{1}{2} \sqrt{\frac{\lambda}{N}} (z+\bar z)})}= \nonumber \\
&&~~~~~~~~ = \mathcal{Z}_N \frac{\partial}{\partial \alpha} \left ( \frac{1}{\mathcal{Z}_N(\alpha)} \int_{[z,\bar z]=0} [d^2 z] e^{-\mbox{\tiny Tr} (z \bar z) + \frac{\alpha}{2}(\mbox{\tiny Tr} z^{\Delta} + \mbox{\tiny Tr} \bar z^{\Delta})} e^{\mbox{\tiny Tr} \log (1+ \frac{1}{t} \, e^{\frac{1}{2} \sqrt{\frac{\lambda}{N}} (z+\bar z)})} \right ) \bigg |_{\, \alpha=0} \nonumber \\
&& ~~~~~~~~\equiv \mathcal{Z}_N \frac{\partial}{\partial \alpha} \left\LA \exp{\left[ \mbox{Tr} \log \left(1+ \frac{1}{t} \, e^{\frac{1}{2} \sqrt{\frac{\lambda}{N}} (z+\bar z)}\right)\right]} \right\RA_{\alpha} \bigg |_{\alpha=0}
\label{nmmantisymm2}
\eea
where we have introduced an $\alpha$-dependent partition function
\bea
\mathcal{Z}_N(\alpha) = \int_{[z,\bar z]=0} [d^2 z] e^{-\mbox{\tiny Tr} (z \bar z) + \frac{\alpha}{2}(\mbox{\tiny Tr} z^{\Delta} + \mbox{\tiny Tr} \bar z^{\Delta})}
\eea
and we have used that $\mathcal{Z}_N(\alpha) = \mathcal{Z}_N + {\cal O} (\alpha^2)$ \footnote{This follows
from the fact that in the matrix model with gaussian potential $\LA \mbox{Tr} z^{\Delta} \RA = \LA \mbox{Tr} \bar z^{\Delta} \RA = 0$.}. The problem is now to evaluate the correlation function (\ref{nmmantisymm2}) in the normal matrix model with the deformed potential
\bea
V(z, \bar z) = -\mbox{Tr} z \bar z + \frac{\alpha}{2} \mbox{Tr} (z^{\Delta} + \bar z^{\Delta}). 
\label{defpot}
\eea
Normal models with potentials of this kind were previously studied in the literature, for a recent account see for example 
\cite{Wiegmann:2003xf}\cite{Zabrodin:2002up}. To solve the model at large $N$, one can as usual go to the eigenvalue basis 
at the expenses of introducing a Vandermonde factor, and determine the eigenvalue density $\rho_{\alpha} (z,\bar z)$ in the continuum limit. The density is found by solving the saddle point equation \footnote{As in \cite{Hartnoll:2006is}, the term $\exp{\left[ \mbox{Tr} \log \left(1+ \frac{1}{t} \, e^{\frac{1}{2} \sqrt{\frac{\lambda}{N}} (z+\bar z)}\right)\right]}$ does not modify the saddle point equation at leading order at large $N$.}
\bea
z-\frac{\Delta \alpha}{2} \bar z^{\Delta-1} = N \int d^2 z' \frac{\rho_{\alpha} (z', \bar z' )}{\bar z - \bar z'}
\label{saddle}
\eea
where the term in the right hand side comes from the Vandermonde factor. Once the density is known, the correlation function 
in (\ref{nmmantisymm2}) becomes
\bea
\left\LA \exp{\left[ \mbox{Tr} \log \left(1+ \frac{1}{t} \, e^{\frac{1}{2} \sqrt{\frac{\lambda}{N}} (z+\bar z)}\right)\right]} \right\RA_{\alpha} 
\rightarrow \exp{\left[ N \int d^2 z \, \rho_{\alpha} (z, \bar z) \log \left(1+ \frac{1}{t} \, e^{\frac{1}{2} \sqrt{\frac{\lambda}{N}} (z+\bar z)}\right)\right]}.\cr &&
\eea 
It is known that for potentials of the kind $V(z, \bar z) = -z \bar z + f(z) + \bar f (\bar z)$, the density is a constant (equal to ${1 \over N \pi}$ in the normalizations we are using here) inside a certain droplet in the complex plane and zero outside. For the gaussian potential, as reviewed previously, the droplet is just a circle of radius $\sqrt{N}$ (to compare 
with eq. (\ref{rhoz}), one has to rescale $z \rightarrow \sqrt{\lambda \over 2 N}z$), 
while the term proportional to $\alpha$ induces a deformation of the circle which preserves its total area (since we do not change the number of eigenvalues). It is not difficult to find the shape of the droplet which solves the saddle point 
equation (\ref{saddle}) at leading order in $\alpha$. It is convenient to work in polar coordinates $z = r e^{i \phi}$. The 
curve which bounds the droplet can then be written at first order as 
\bea
r(\phi) = \sqrt{N} \big( 1 + \alpha \, f(\phi) \big ).
\eea 
Clearly $f(\phi)$ has to be periodic, and may be written as
\bea
f(\phi) = \sum_{n=1}^{\infty} a_n \, \cos n \phi 
\label{fourier}
\eea
where only cosines appear because of the symmetry of the potential (\ref{defpot}) under $z \leftrightarrow \bar z$, and the mode with $n=0$ is excluded by requiring the area to be preserved. The saddle point equation now reads
\bea
r \, e^{i \phi} - \frac{\Delta \alpha}{2} r^{\Delta-1} \, e^{-i (\Delta-1) \phi} = \frac{1}{\pi} \int_{0}^{2 \pi} d \phi' \int_{0}^{\sqrt{N} (1 + \alpha \, f(\phi'))} dr' \, \frac{r'}{r\, e^{-i \phi} - r' \, e^{-i \phi'}}.
\eea 
Expanding the integral at first order in $\alpha$ and plugging in the Fourier expansion (\ref{fourier}), we see that this equation is satisfied if $a_{\Delta} = N^{\Delta/2 -1} \, \frac{\Delta}{2}$ and all other $a_n$ vanish, so we find that the shape of the deformed droplet is given by the curve 
\bea
r(\phi) = \sqrt{N} \left( 1 + \frac{\alpha}{2} \, \Delta \, N^{\Delta/2 -1} \cos \Delta \phi \right).
\label{defdrop}
\eea 

Before moving on to compute (\ref{nmmantisymm2}), we can check the validity of the method by applying it to the computation 
of the correlator (\ref{wo}) when the Wilson loop is in the fundamental representation. In this case, following the same steps as above, in the large $N$ limit we arrive at
\bea
\LA W_{\square}\mathcal{O}_\D \RA &=& \frac{2^{\Delta/2}}{\sqrt{\Delta} N^{\Delta/2}} \frac{\partial}{\partial \alpha} 
\int d^2 z \, \rho_{\alpha} (z, \bar z) \, e^{\frac{1}{2} \sqrt{\frac{\lambda}{N}}(z+\bar z)} \Big |_{\alpha=0} \nonumber \\
&=& \frac{2^{\Delta/2}}{\sqrt{\Delta} N^{\Delta/2}} \frac{\partial}{\partial \alpha} 
\frac{1}{N \pi} \int_0^{2 \pi} d\phi \int_{0}^{\sqrt{N}(1 + \frac{\alpha}{2} \, \Delta \, N^{\Delta/2 -1} \cos \Delta \phi)} dr\,r \, e^{\sqrt {\frac{\lambda}{N}} r \cos \phi} \bigg |_{\alpha=0} \nonumber \\
&=&  \frac{2^{\Delta/2} \sqrt{\Delta}}{N} \frac{1}{2 \pi} \int_{0}^{2 \pi} d \phi \, e^{\sqrt{\lambda} \cos \phi} \cos \Delta \phi =  \frac{2^{\Delta/2} \sqrt{\Delta}}{N} I_{\Delta} (\sqrt{\lambda})
\label{fundam}  
\eea  
which is the result first found in \cite{Semenoff:2001xp} and the correct large $N$ limit of the exact formula (\ref{mm}) (with $k=1$).

Going back to the antisymmetric representation, we have to evaluate 
\bea
&& \oint \frac{dt}{2 \pi i} t^{k-1} \frac{\partial}{\partial \alpha} \exp{\left[ N \int d^2 z \, \rho_{\alpha} (z, \bar z) \log \left(1+\frac{1}{t} \, e^{\frac{1}{2} \sqrt{\frac{\lambda}{N}} (z+\bar z)}\right)\right]} \bigg |_{\alpha=0} \nonumber \\ 
&& = \oint \frac{dt}{2 \pi i} t^{k-1} \frac{\partial}{\partial \alpha} \left[ N \int d^2 z \, \rho_{\alpha} (z, \bar z) \log \left(1+\frac{1}{t} \, e^{\frac{1}{2} \sqrt{\frac{\lambda}{N}} (z+\bar z)}\right)\right] \bigg |_{\alpha=0} \times \nonumber \\ 
&& \qquad \qquad \qquad  \times \, \exp{\left[ N \int d^2 z \, \rho_{0} (z, \bar z) \log \left(1+\frac{1}{t} \, e^{\frac{1}{2} \sqrt{\frac{\lambda}{N}} (z+\bar z)}\right)\right]}
\label{antisymm}
\eea
where in the last line $\rho_0$ is just the circular droplet density. Since the exponent is independent of $\alpha$, the $t$ 
integral can be evaluated in the supergravity limit of large $\lambda$ exactly as in \cite{Hartnoll:2006is}: we first make a change of variables $t=e^{\sqrt{\lambda} w}$, then the saddle point of the exponent is found to be
\bea
w_{\star} = \cos \theta_k
\eea
where $\theta_k$ is defined as in eq. (\ref{thetak}). The exponent in (\ref{antisymm}) gives a term proportional to the 
expectation value of the Wilson loop, while the prefactor is evaluated at the saddle point. After dividing by $\LA W_{A_k} \RA$, the OPE coefficient can then be obtained as
\bea
&& \frac{\LA W_{A_k} \, {\cal O}_{\Delta} \RA}{\LA W_{A_k} \RA} = \frac{2^{\Delta/2}}{\sqrt{\Delta} N^{\Delta/2}} 
\frac{\partial}{\partial \alpha} \left[ N \int d^2 z \, \rho_{\alpha} (z, \bar z) \log \left(1+ \, e^{\frac{1}{2} \sqrt{\lambda} \left( \frac{z+\bar z}{\sqrt{N}}-2 \cos \theta_k\right)}\right)\right] \bigg |_{\alpha=0} \nonumber \\
&& \simeq \frac{2^{\Delta/2}N\sqrt{\lambda}}{\sqrt{\Delta} N^{\Delta/2}} 
\frac{\partial}{\partial \alpha} \frac{2}{\pi} \int_0^{\theta_k} d\phi \int_{\frac{\cos \theta_k}{\cos \phi}}^
{1 + \frac{\alpha}{2} \, \Delta \, N^{\Delta/2 -1} \cos \Delta \phi} dr \, r (r \cos \phi -\cos \theta_k) \Big |_{\alpha=0}
\label{antisymmcoef}
\eea
where the lower limit in the $r$ integral comes from the fact that in the large $\lambda$ limit the integral has support only in the region $r \cos \phi  \ge  \cos \theta_k$ \footnote{The upper limit in the $\phi$ integral is rigorously $\theta_k + 
{\cal O} (\alpha)$, but it is easy to see that the correction does not contribute at first order in $\alpha$.}. After 
doing the derivative, (\ref{antisymmcoef}) gives the final result
\bea
\frac{\LA W_{A_k} \, {\cal O}_{\Delta} \RA}{\LA W_{A_k} \RA} = \frac{2^{\Delta/2} \sqrt{\Delta \lambda}}{\pi} 
\int_{0}^{\theta_k} \, d\phi \, \cos \Delta \phi \, (\cos \phi - \cos \theta_k).
\label{final0}
\eea     
Remarkably, the integral in (\ref{final0}) precisely reproduces the Gegenbauer polynomials arising in the bulk computation, and the final result is 
\bea
\frac{\LA W_{A_k} \, {\cal O}_{\Delta} \RA}{\LA W_{A_k} \RA}  = {2^{\Delta / 2} \over 3 \pi} \sqrt{\Delta \lambda} \sin^3
\theta_k \,
 {6 (\Delta -2) !  \over (\Delta + 1)!} \,  C_{\Delta-2}^{(2)} (\cos \theta_k)
\eea
which exactly matches the $D5$ computation of the OPE coefficient.

\section{Conclusion}
\setcounter{equation}{0}

In this paper we computed the correlation function between a higher
rank Wilson loop and a chiral primary operator  in the fundamental representation using branes with
electric fluxes. Following the proposal of \cite{Drukker:2005kx}\cite{Yamaguchi:2006tq}\cite{Gomis:2006sb}, we considered a D$3_k$ brane for the rank $k$ symmetric
case and a D$5_k$ brane for the antisymmetric one. We then checked
our results with the normal matrix model discussed in
\cite{Okuyama:2006jc} finding perfect agreement in both cases.

We focussed on chiral primary operators but it should not be difficult to
extend our computation to operators corresponding to other supergravity modes. For example, the KK modes of the dilaton are necessary to compute correlation functions of Wilson loops and $\mbox{Tr}\,\Phi^\D F^2_+$.

It would be worthwhile to study more general representations of both
the Wilson loop and the chiral primary operator. A particularly
interesting issue to address would be understanding from our brane
picture the selection rule found in \cite{Okuyama:2006jc}: for
Wilson loops in the rank $k$ antisymmetric representation the only
non vanishing correlators involve chiral primaries with traces over
Young diagrams with at most $k$ hooks.
Another direction to pursue may be considering the correlation function between
higher dimensional Wilson loop and a chiral primary operator with $\D\sim N$. In
the bulk this would require to study the bulk-to-bulk exchange of supergravity degrees of
freedom between the electric branes describing the Wilson loop and the (dual) giant gravitons associated with the chiral primary.

\subsubsection*{Acknowledgements}

We would like to express our gratitude to the participants and
organizers of the 4th Simons Workshop in Mathematics and Physics at
Stony Brook for creating a stimulating environment from which we
benefited greatly. We acknowledge partial financial support through
NSF award PHY-0354776.

\appendix
\section{Spherical harmonics and orthogonal polynomials}

In this appendix we collect some facts about spherical harmonics
and orthogonal polynomials we have used in the paper. We follow
the treatments of \cite{Higuchi:1986wu} and \cite{szego}.

Spherical harmonics in $d$ dimensions are eigenfunctions of the
Laplacian on the unit $d$-sphere \bea
\nabla^2_{(d)}Y^{I}(\Omega)=\lambda
Y^{I}(\Omega)\label{eigenproblem}\eea where the Laplacian is \bea
\nabla^2_{(d)}=\frac{1}{\sqrt{\det g}}\,\p_i\,
\sqrt{\det g}\,g^{ij}\,\p_j\label{laplacianSn}\eea
with the metric given by
$g_{ij}=\mbox{diag}(1,\sin^2\theta_d(1,\sin^2\theta_{d-1}(\ldots)))$.
 The integer multi-index $I=(l_d,\ldots\, ,l_1)$ satisfies \be
l_d\geq l_{d-1}\geq\cdots\geq l_2\geq |l_1|. \ee The general
solution to eq. (\ref{eigenproblem}) is \be Y^{l_d,\ldots,
l_1}(\theta_d,\ldots,\theta_1)={e^{i\,l_1\theta_1} \over \sqrt{2
\pi}}\prod_{n=2}^{d}\,{}_n\bar{P}_{\,l_n}^{\, l_{n-1}}(\theta_n) \ee
where we defined \be {}_n\bar{P}_{\,L}^{\, l}(\theta)= {}_nc_{\,L}^{\,
l}(\sin{\theta})^{-(n-2)/2}P_{\,L+(n-2)/2}^{\,-(l+(n-2)/2)}(\cos{\theta}).\ee
In this expression $P^{\,-m}_n(x)$ is the Legendre
function of the first kind and the constant \be {}_nc_{\,L}^{
\,l}=\left[{(2L+n- 1)(L+l+n-2)!\over{2 (L-l)!}}\right]^{1/2} \ee  is
chosen to ensure
the orthonormalization condition 
\be \int \mu(\Omega_{d})\,\,Y^I
Y^{I'}=\delta^{I I'} \ee where $\mu(\Omega_{d})$ is the measure over $S^d$. The integration over $S^{d-1}$ selects
only $SO(d)$ invariant harmonics
\be \int \m(\Omega_{d-1})\sum_I Y^I=\sum_{l_d}\,Y^{l_d,\,0,\ldots,0}. \ee

The eigenvalue $\lambda$ depends only on $l_d\equiv\D$ because of
the $O(d+1)$ symmetry of the problem and it can be found by
studying the action of the Laplacian on $SO(d-1)$ invariant
spherical harmonics \bea
\left(\frac{1}{\sin^{d-1}\theta_d}\frac{\p}{\p \theta_d}\,
\sin^{d-1}\theta_d\,\frac{\p}{\p \theta_d}\right)Y^{\D,\,
0}(\Omega)=\lambda_{\D}Y^{\D,
\,0}(\Omega).\label{eigenproblem1}\eea After the change of
variable $x=\cos\theta_d$, this is recognized to be the Gegenbauer
equation \bea \left((1-x^2)\frac{\p^2}{\p x^2}\, -d\, x\frac{\p}{\p
x}\right)Y^{\D,\, 0}(x)=\lambda_{\D}Y^{\D,\,
0}(x).\label{eigenproblem2}\eea The solution to this equation is
\bea \lambda_\D=-\D(\D+d-1)\, ,~~~~~~~~~~~Y^{\D,\,
0}(x)=\mathcal{N}_\D\, C_\D^{\left(\frac{d-1}{2}\right)}(x)\eea
where $C_\D^{\left(\frac{d-1}{2}\right)}$ are Gegenbauer
polynomials and the constant $\mathcal{N}_\D$ can be obtained from
the orthonormality of the $Y^{\D,\,0}$'s \bea
\mathcal{N}_\D=\left[\frac{\D!(2\D+d-1)\left[\Gamma\left(\frac{d-1}{2}\right)\right]^2
\Gamma\left(\frac{d}{2}\right)}{2^{4-d}\pi^{\frac{d+2}{2}}\Gamma(\D+d-1)}
\right]^{1/2}.\eea

The Gegenbauer polynomials $C^{(\lambda)}_\D(x)$ are a
generalization of the Legendre polynomials and can be obtained
from the following generating function \bea
\frac{1}{(1-2xt+t^2)^\lambda}=\sum_{\D=0}^{\infty}C^{(\lambda)}_\D(x)\,
t^\D.\eea We list the first few of them \bea &&
C^{(\lambda)}_0(x)=1\cr && C^{(\lambda)}_1(x)=2\lambda x\cr &&
C^{(\lambda)}_2(x)=-\lambda+2\lambda(1+\lambda)x^2\cr &&
C^{(\lambda)}_3(x)=-2\lambda(1+\lambda)x+\frac{4}{3}\lambda(1+\lambda)
(2+\lambda)x^3.\eea They satisfy the normalization condition \bea
\int_{-1}^1 dx
(1-x^2)^{\lambda-1/2}\left[C^{(\lambda)}_\D\right]^2
=2^{1-2\lambda}\pi\frac{\Gamma(\D+2\lambda)}{(\D+\lambda)\Gamma^2(\lambda)\Gamma(\D+1)}\eea
for $\lambda > -1/2$.

In the paper we have used the following formula for the derivative
of a Gegenbauer polynomial
 \bea (1-x^2)\partial_x
C_\D^{(\lambda)}(x) = -\D\, x\, C_\D^{(\lambda)}(x)+
(\D+2\lambda-1) C_{\D-1}^{(\lambda)}(x) \label{gegenderivapp} \eea
and the following recurrence relation \bea \D\,
C_\D^{(\lambda)}(x) = 2 (\D+\lambda-1) x\,
C_{\D-1}^{(\lambda)}(x)- (\D+2\lambda-2) C_{\D-2}^{(\lambda)}(x).
\label{gegenrecurr} \eea We have also used that \bea
C_{\D}^{\left(\frac{d-1}{2}\right)}(1)=\frac{(\D+d-2)!}{\D!(d-2)!}.\label{C1}\eea



\end{document}